\documentclass[runningheads]{llncs}

\usepackage{subcaption}
\usepackage{comment}
\usepackage{amsmath}
\usepackage{amssymb}
\usepackage{amsfonts}
\usepackage{graphicx}
\usepackage{color}
\usepackage{multicol}
\usepackage{multirow}
\usepackage[ruled,noline]{algorithm2e}
\usepackage{comment}
\usepackage{booktabs}
\usepackage{subcaption}
\usepackage{lipsum}
\newcommand{\bumpup}{\vspace*{-2.0ex}}

\usepackage{booktabs}

\usepackage{graphicx}
\begin{document}
	\title{Integrating cross-modality hallucinated MRI with CT to aid mediastinal lung tumor segmentation} 
	%
	%
	
	\author{Jiang Jue\inst{1}
		\and  Hu Jason \inst{1} 
		\and Tyagi Neelam\inst{1}
		\and Rimner Andreas\inst{2}
		\and Berry L. Sean\inst{1}
		\and Deasy O. Joseph\inst{1}
		\and Veeraraghavan Harini\inst{1} \thanks{Email: veerarah@mskcc.org:This paper has been accepted by MICCAI 2019}}

	\institute{Medical Physics, Memorial Sloan Kettering Cancer Center \and Radiation Oncology, Memorial Sloan Kettering Cancer Center}

	\authorrunning{Jiang Jue et al.}
	%
	
	\maketitle              
	\begin{abstract}
		Lung tumors, especially those located close to or surrounded by soft tissues like the mediastinum, are difficult to segment due to the low soft tissue contrast on computed tomography images. Magnetic resonance images contain superior soft-tissue contrast information that can be leveraged if both  modalities  were  available  for  training.  Therefore, we developed a cross-modality educed learning approach where MR information that is educed from CT is used to hallucinate MRI and improve CT segmentation. Our approach, called cross-modality educed deep learning segmentation (CMEDL) combines CT and pseudo MR produced from CT by aligning their features to obtain segmentation on CT. Features computed in the last two layers of parallelly  trained CT and MR segmentation networks are aligned. We implemented this approach on U-net and dense fully convolutional networks (dense-FCN). Our networks were trained on unrelated cohorts from open-source the Cancer Imaging Archive CT images (N=377), an internal archive T2-weighted MR (N=81), and evaluated using separate validation (N=304) and testing (N=333) CT-delineated tumors.  Our approach using both networks were significantly more accurate (U-net $P <0.001$; denseFCN $P <0.001$) than CT-only networks and achieved an accuracy (Dice similarity coefficient) of 0.71$\pm$0.15 (U-net), 0.74$\pm$0.12 (denseFCN) on validation and 0.72$\pm$0.14 (U-net), 0.73$\pm$0.12 (denseFCN) on  the  testing sets.  Our novel approach demonstrated that educing cross-modality information through learned priors enhances CT segmentation performance.

	\end{abstract}

	\section{Introduction}
\begin{figure*}
		\begin{center}
			\includegraphics[width=1\columnwidth,scale=1]{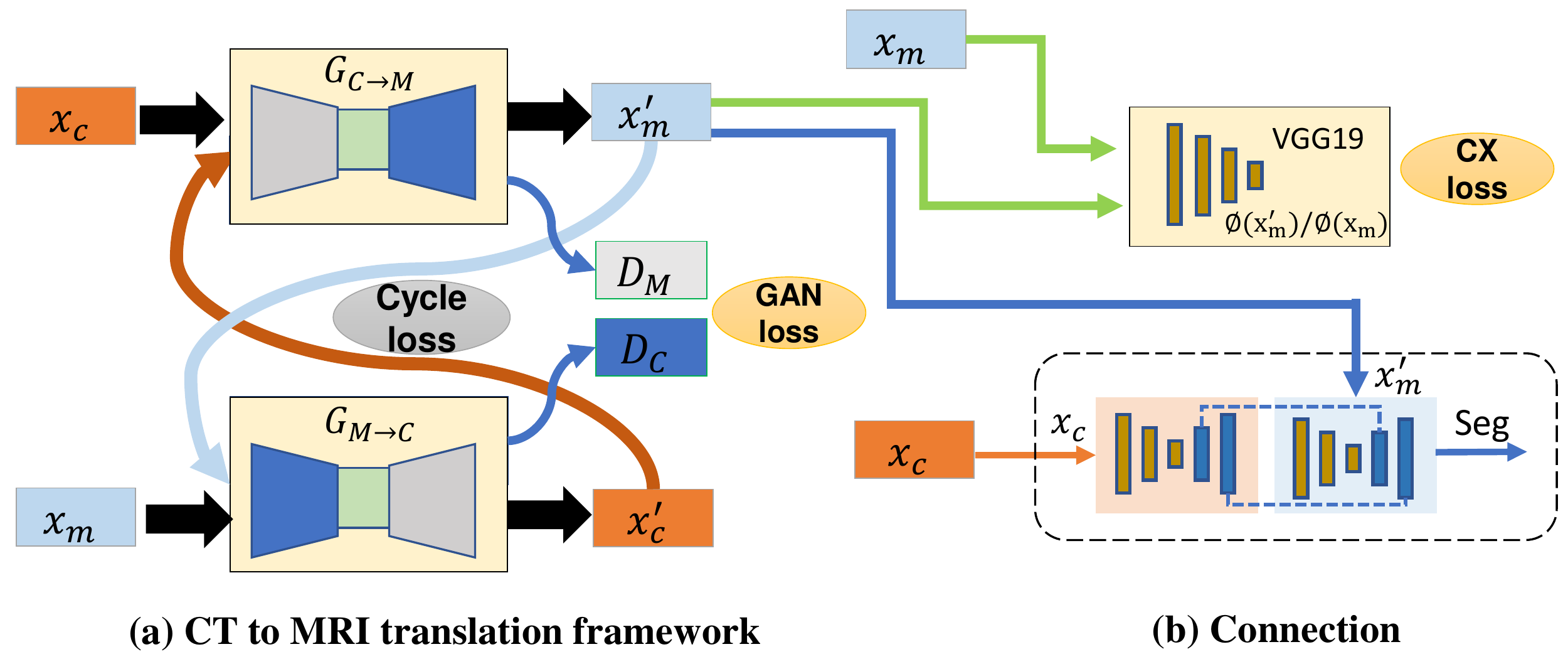}
			\vspace{-0.05cm}\setlength{\belowcaptionskip}{-0.4cm}\setlength{\abovecaptionskip}{0.08cm}\caption{\label{fig:methods} \small Overview of the comparison of different combinations. $x_{c}$ is the CT image; $x_{m}$ is the MRI image;$G_{C \rightarrow M}$ and $G_{M \rightarrow C}$ are the CT and MRI transfer networks; $x_{m}^{'}$ is the translated MRI image from $x_{c}$; $x_{c}^{'}$ is the translated MRI image from $x_{m}$.}
		\end{center}
	\end{figure*}

	\bumpup
	Precision medical treatments including image-guided radiotherapy require accurate target tumor segmentation~\cite{njeh2008tumor}. Computed tomography (CT), the standard-of-care imaging modality lacks sufficient soft-tissue contrast, which makes visualizing tumor boundaries difficult, especially for those that are adjacent to soft-tissue structures. With the advent of new MRI simulator technologies, radiation oncologists can delineate target structures on MRI acquired in simulation position, which then have to be transferred using image registration to the planning CTs acquired at a different time in treatment position for radiation therapy planning~\cite{devic2012}. Image registration itself is prone to errors and thus accurate segmentation on CT itself is more desirable for improving accuracy of clinical radiation treatment margins. More importantly, driven by the lack of simultaneously acquired CT and MR scans, current methods are restricted to CT alone. Therefore, we developed a novel approach, called cross-modality educed deep learning (CMEDL), that uses unpaired cross-domain adaptation between unrelated CT and MR datasets to hallucinate MR-like images or pseudo MR (pMR) from CT scans. The pMR image is combined with CT image to regularize training of a CT segmentation network. This is accomplished by aligning the features of the CT with the pMR features during training (Figure.~\ref{fig:methods}). 
	\\
	Ours is not a method for data augmentation using cross-domain adaptation~\cite{nie2017medical,chartsias2017adversarial,jiang2018tumor}. Our work is also unlike methods that seek to reduce the datashift differences between same imaging modalities~\cite{zhu2017unpaired,long2017deep,kamnitsas2017efficient}. Instead, our goal in this work is to maximize the segmentation performance in a single less informative imaging modality, namely, CT using learned information modeling the latent tissue relationships with a more informative modality, namely MRI. The key insight here is that the features dismissed as uninterpretable on CT can provide inference information when learning proceeds from a more informative modality such as MRI. 
	\\
	Our approach is most similar in its goal to compute shared representations for improving segmentations as in the work by~\cite{valindria2018}, where several shared representations between CT and MRI were constructed using fully convolutional networks. Our approach, that is based on GANs for cross-modality learning, shares some similarities to~\cite{cai2018towards} that also used a GAN as a backbone framework, and implemented dual networks for performing segmentations on both CT and MRI. However, our approach substantially differs from prior works in its use of the cross-modality tissue relations as priors to improve inference on the less informative source (or CT) domain. Though applied to segmenting lung tumors, this method is generally applicable to other structures and imaging modalities.
	\\
	Our contributions in this work are as follows: (i) first, we developed a novel approach to generate segmentation on CT by leveraging more informative MRI through cross-modality priors. (ii) second, we implemented this approach on two different segmentation networks to study feasibility of segmenting lung tumors located in the mediastinum, an area where there is diminished contrast between tumor and the surrounding soft-tissue. (iii) third, we evaluated our approach on a large dataset of 637 tumors.
	
	\section{Methods} 
	\bumpup
	We use a supervised cross-modality and CT segmentation approach with a reasonably large number of expert segmented CT scans ($X_{CT}, y_{CT}$) and a few MR scans with expert segmentation ($\{X_{MR} , y_{MR}\}$, where, $N_{X_{MR}}$ $\ll$ $N_{X_{CT}}$). The cross-modality educed deep learning (CMEDL) segmentation consists of two sub-networks that are optimized alternatively. The first sub-network (Figure.~\ref{fig:methods} A) generates a pMR image given a CT image. The second sub-network (Figure.~\ref{fig:methods} B), trains its CT segmentation network constrained using features from another network trained using pMRI. The alternative optimization enables the approach to regularize both the segmentation and pMR generation, such that the pMR is specifically tuned to increase segmentation accuracy. In other words, pMR acts as an informative regularizer for CT segmentation, while the gradients of segmentation errors serve to constrain the generated pMR images. 
	
	\subsection{Cross-domain adaptation for hallucinating pseudo MR images:}
	A pair of conditional GANs\cite{goodfellow2014generative} are trained with unpaired CT and T2-weighted (T2w) MR images arising from different sets of patients. The first GAN transforms CT into a pseudo MR (pMR) image ($G_{C \rightarrow M}$) while the second, transforms a MR image into its corresponding pseudo CT (pCT) ($G_{M \rightarrow C}$) image. The GANs are optimized using the standard adversarial loss ($L_{adv} = L_{adv}^{CT}+L_{adv}^{MR}$) and cycle consistency losses ($L_{cycl} = L_{cycl}^{CT} + L_{cycl}^{MR}$). In addition, we employed a contextual loss that was introduced for real-world images~\cite{mechrez2018contextual} in order to handle learning from image sets lacking spatial correspondence. The contextual loss facilitates such transformations by treating images as collection of features and computing a global similarity between all pairs of features between the two images ($\{g_{j \in N}, m_{i \in M}\}$) used in computing domain adaptation. The contextual similarity is expressed as:
	\begin{equation}
	CX(g,m) = \frac{1}{N}\sum_{j} \underset{i}max CX(g_{j},m_{i}),
	\end{equation} 
	where, $N$ corresponds to the number of features. The contextual similarity is computed by normalizing the inverse of cosine distances between the features in the two images as described in\cite{mechrez2018contextual}. The contextual loss is computed as:
	\begin{equation}
	L_{cx} = -log(CX(f(G(X_{CT})), f(X_{MR})).
	\end{equation}
	The total loss for the cross-modality adaptation is then expressed as the summation of all the aforementioned losses. The pMR generated from this step is passed as an additional input for training the CT segmentation network. 
	
	\subsection{Segmentation combining CT with pMR images}
	\bumpup
	Our approach for combining the CT with pMR images 
	uses the idea of only matching information that is highly predictable from each other. This usually corresponds to the features closest to the output as the two images are supposed to produce identical segmentation. Therefore, the features computed from the last two layers of CT and pMR segmentation networks are matched by minimizing the squared difference or the L2 loss between them. This is expressed as below.
	\textcolor{black}{\begin{equation}
		\setlength{\abovedisplayskip}{1pt}
		\setlength{\belowdisplayskip}{1pt}
		\begin{split}
		L_{seg}  & = \mathbb{E}_{x_{c}\sim X_{CT}}[-log P(S_{MR}(G_{CT\rightarrow MR}(x_{c}))) -log P(S_{CT}(x_{c}))] + \\ &  \|\phi{_{CT}}(x_{c})-\phi{_{MR}}(G_{CT\rightarrow MR}(x_{c}))||^{2}_{F}, 
		\label{eqn:C5}
		\end{split}
		\end{equation}}\\
	where $S_{CT}, S_{MR}$ are the segmentation networks trained using the CT and pMR images, $\phi_{CT}, \phi_{MR}$ are the features computed from these networks, and $G_{CT \rightarrow MR}$ is the cross-modality network used to compute the pMR image, and $F$ stands  for Frobenius norm. 
	
	The total loss computed from the cross-modality adaptation and the segmentation networks is expressed as:
	\begin{equation}
	\textrm{Loss} = L_{adv} + \lambda_{cyc} L_{cyc} + \lambda_{cx} L_{CX} + \lambda_{seg} L_{seg},  
	\end{equation}
	where $\lambda_{cyc}$, $\lambda_{cx}$ and $\lambda_{seg}$ are the weighting coefficients for each loss. During training, we alternatively update the cross-domain adaptation network and the segmentation network with the following gradients, $-\Delta_{\theta_{G}}(L_{adv}+ \lambda_{cyc}{L_{cyc}}+ \lambda_{c}{L_{c}}+ \lambda_{cx}L_{cx})$, $-\Delta_{\theta_{D}}(L_{adv})$ and $-\Delta_{\theta_{seg}}L_{seg}$. More concretely, the segmentation network is fixed when updating the cross-modality translation and vice versa in each iteration.

	\subsection{Segmentation architecture:}
	We implemented the U-net\cite{ronneberger2015u} and dense fully convolutional networks (denseFCN)~\cite{jegou2017one} to evaluate the feasibility of combining hallucinated MR for improving CT segmentation accuracy. These networks are briefly described below. 
	\bumpup
	\begin{enumerate}
		\item \textbf{U-net \/}\rm was modified using batch normalization after each convolution filter in order to standardize the features computed at the different layers. 
		
		\item \textbf{Fully Convolutional DenseNets \/}\rm (Dense-FCN) that is based on \cite{jegou2017one}, uses dense feature maps computed using a sequence of dense feature blocks and concatenated with feature maps from previous computations through residual connections. Specifically, a dense feature block is produced by iterative summation of previous feature maps within that block. As features computed from all image resolutions starting from the image resolution to the lowest resolution are iteratively concatenated, features at all levels are utilized. This in turn facilitates an implicit dense supervision to stabilize training.
	\end{enumerate} 
	\bumpup
	\subsection{Implementation and training}
	All networks were implemented using the Pytorch\cite{paszke2017automatic} library and trained end to end on Tesla V100 with 16 GB memory and a batch size of 2. The ADAM algorithm \cite{kingma2014adam} with an initial learning rate of 1e-4 was used during training. The segmentation networks were trained with a learning rate of 2e-4. We set $\lambda_{adv}$=10, $\lambda_{cx}$=1, $\lambda_{cyc}$=1 and $\lambda_{seg}$=5. For the contextual loss, we use the convolution filters after the Con7, Conv8 and Conv9 due to memory limitations.
	
	\section{Datasets and evaluation}
	\bumpup
	We used patients obtained from three different cohorts consisting of (a) the Cancer Imaging Archive (TCIA)\cite{clark2013cancer} with  non-small cell lung cancers (NSCLC)~\cite{aerts2014decoding} consisting of 377 patients (training), (b) 81 longitudinal T2-weighted MR scans (scanned on Philips 3T Ingenia) from 21 patients treated with radiation therapy, and (training) (c) 637 contrast-enhanced tumors treated with immunotherapy at our institution for validation (N=304) and testing (N=333) such that different sets of patients were used for validation and testing. \textcolor{black}{Early stopping was used during the training to prevent overfitting and the best model selected using validation set was used for testing. Identical CT datasets were used in both CT only and CMEDL approach for  equitable comparisons.} Expert segmentations were available on all scans. 
	\\
	The segmentation accuracies were evaluated using Dice similarity coefficient (DSC) and \textcolor{black}{Hausdorff} distance at $95^{th}$ percentile (HD95) as recommended in~\cite{menze2015multimodal}. In addition, we computed the detection rate for the tumors where tumors with at least 50\% DSC overlap with expert segmentations were considered as detected. 
	\section{Results}
	\bumpup
	\subsection{Tumor detection rate}
	\bumpup
	Our method achieved the most accurate detection using both U-net and DenseFCN methods for validation and test sets. In comparison the CT-only method resulted in much lower detection rates for both networks (
	Table~\ref{tab:detectionrate_andseg}).
	\bumpup
	\begin{table*} 
		\centering{\caption{Detection and segmentation accuracy using the two networks.} 
			\label{tab:detectionrate_andseg} 
			\centering
			\scriptsize
			\begin{tabular}{|c|c|c|c|c|c|c|} 
				\hline 
				\multicolumn{1}{|c|}{ }& \multicolumn{3}{|c|}{Validation} & \multicolumn{3}{|c|}{Test}\\ 
				\hline 
				{  Method  }  & {Detection rate} & {  DSC  }  & {  HD95 $mm$ }& {Detection rate} & {  DSC  }& {  HD95 $mm$  }\\
				\hline 
				\multirow{1}{*}{U-net CT} & 80\% & {0.67$\pm$0.18} & {7.44$\pm$7.18 } & 79\% & { 0.68$\pm0.17$ }  & { 9.35$\pm7.08$ }\\
				\hline			
				\multirow{1}{*}{DenseFCN CT} & 77\% &  { 0.70$\pm$0.15 } & { 7.25$\pm$6.71   } &  75\% & { 0.68$\pm$0.16  } & { 9.34$\pm$9.68   }\\
				\hline	
				\multirow{1}{*}{U-net CMEDL} & 85\% &  { 0.71$\pm$0.15 } & { 6.57$\pm$7.15   }& 85\% & { 0.72$\pm0.14$ }& { 8.22$\pm6.89$ }\\
				\hline
				\multirow{1}{*}{DenseFCN CMEDL } & 84\% & { 0.74$\pm$0.12}& { 5.89$\pm$5.87}& 84\% & {0.73$\pm0.12$}& { 7.19$\pm8.55$}\\
				\hline
		\end{tabular}} 
	\end{table*}
\begin{figure}
	\begin{center}
		\includegraphics[width=0.8\columnwidth,scale=0.8]{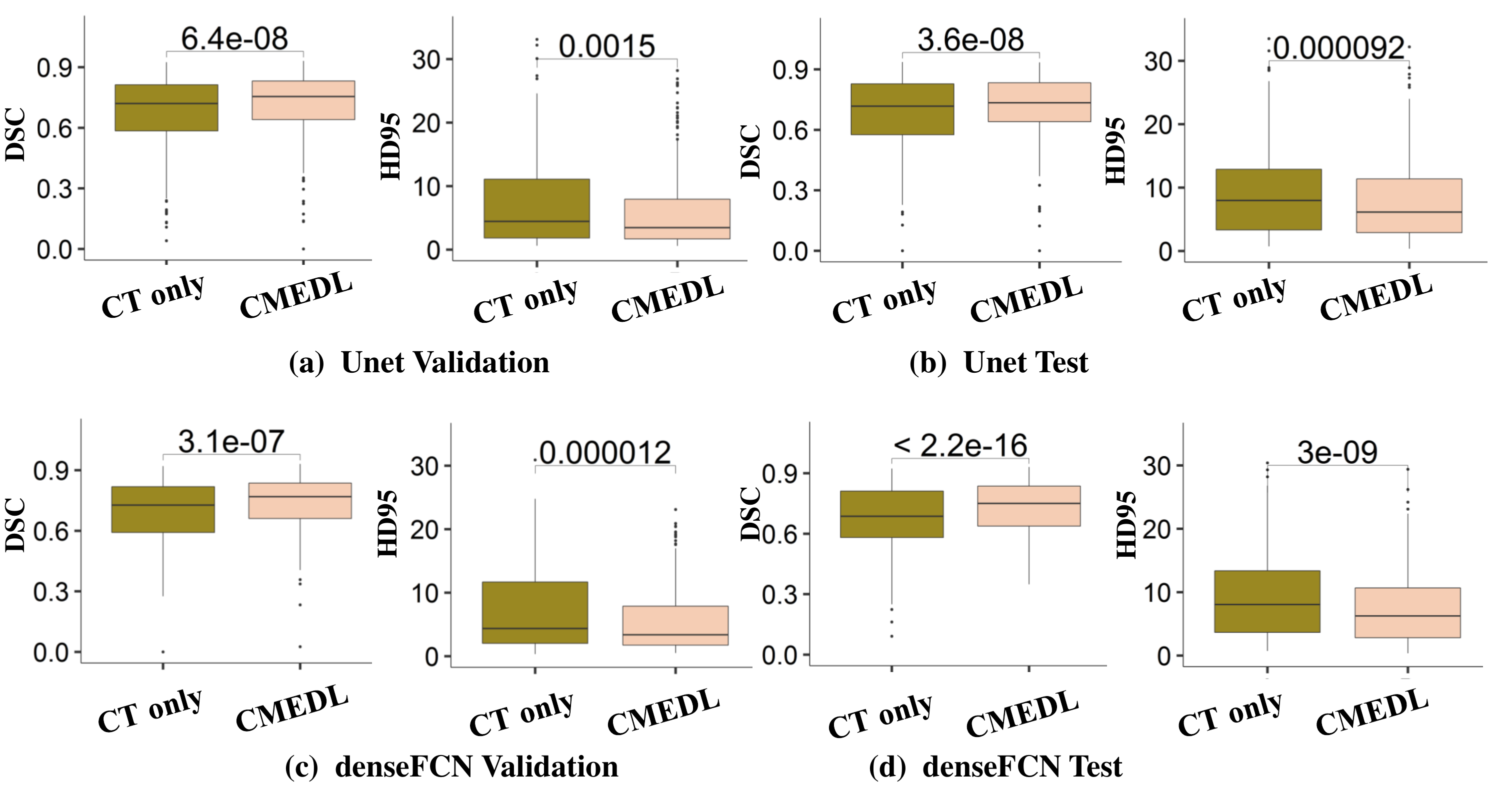}
		\vspace{-0.05cm}\setlength{\belowcaptionskip}{-0.4cm}\setlength{\abovecaptionskip}{0.08cm}\caption{\label{fig:box} \small Box plots comparing CT-only and CMEDL-based networks.}
	\end{center}
\end{figure} 
	\subsection{Segmentation accuracies}
	
	The CMEDL approach resulted in more accurate segmentations than CT-only segmentations (see Table~\ref{tab:detectionrate_andseg}). In addition, both of the U-net and denseFCN networks trained using CMEDL approach were significantly more accurate than CT only segmentations when evaluated with both DSC ($P < 0.001$) and HD95 ($P < 0.001$) metrics. \textcolor{black}{Figure~\ref{fig:box} shows the box plots for the validation and test sets using the two metrics and the two networks. P-values computed using paired Wilcoxon two-sided tests are also shown.} 
	
	\subsection{Visual comparisons}
\begin{figure}
		\begin{center}
			\includegraphics[width=0.8\columnwidth,scale=0.8]{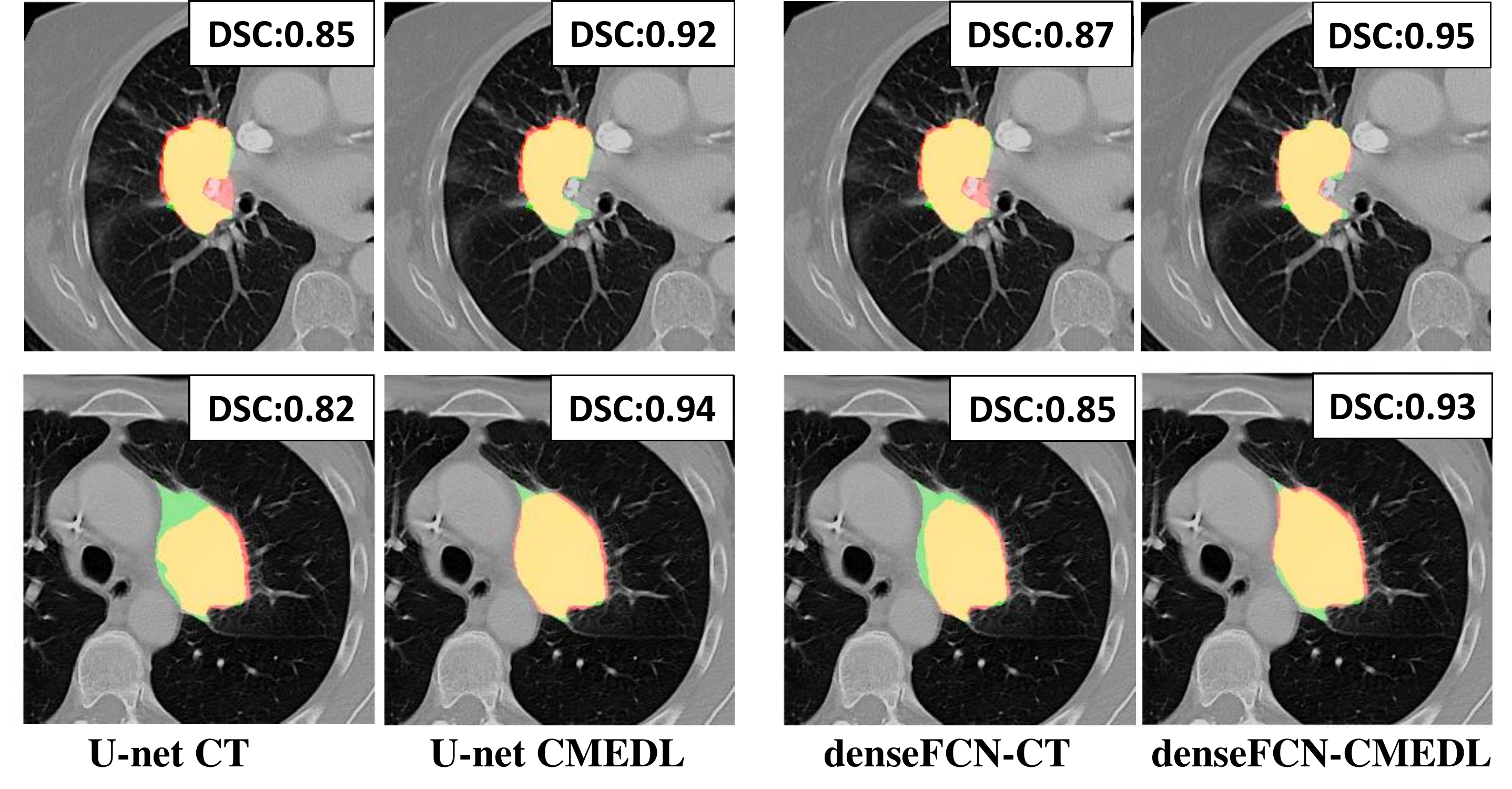}
			\vspace{-0.05cm}\setlength{\belowcaptionskip}{-0.4cm}\setlength{\abovecaptionskip}{0.08cm}\caption{\label{fig:seg_overlay} \small  Representative segmentations produced using CT-only and CMEDL-based segmentations for U-net and DenseFCN networks. The Dice similarity coefficient (DSC) is also shown for each method. Red corresponds to algorithm, green to expert and yellow to overlap between algorithm and expert.}
		\end{center}
	\end{figure}

\begin{figure}
		\begin{center}
			\includegraphics[width=0.8\columnwidth,scale=0.8]{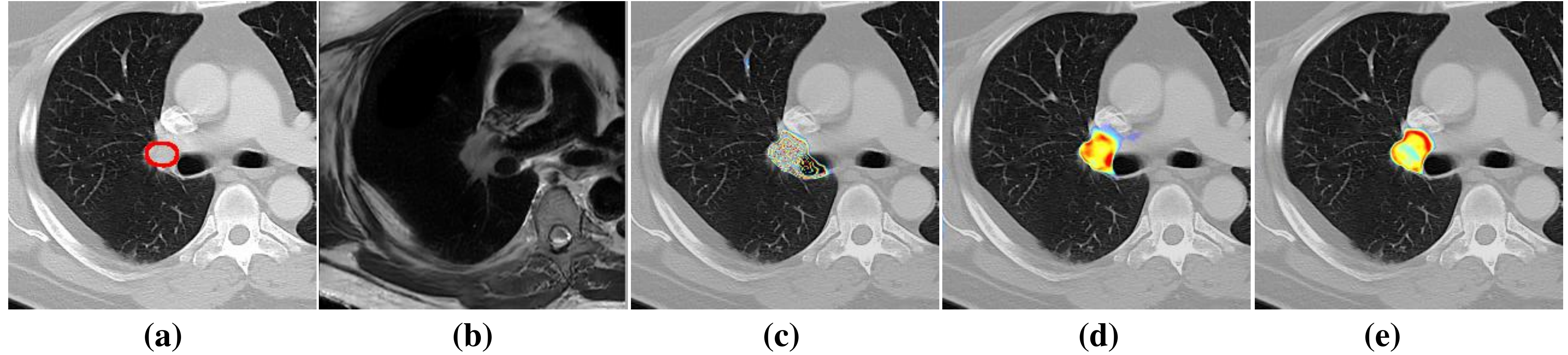}
			\vspace{-0.05cm}\setlength{\belowcaptionskip}{-0.4cm}\setlength{\abovecaptionskip}{0.08cm}\caption{\label{fig:featuremap} \small  Feature map activations from the 21 channel of last layer of Unet. (a) the original CT (b) the translated pMRI (c) activation from CT only (d) activation from pMRI (e) activation from CMEDL}
		\end{center}
	\end{figure}

	Figure~\ref{fig:seg_overlay} shows visual segmentation results produced by the different networks for representative cases when trained using CT-only and with the CMEDL approach. As seen, in both networks, the CMEDL method closely follows the expert-segmentation that is missed using CT-only networks. Figure~\ref{fig:featuremap} shows the feature map activations produced using U-net CT only and with Unet CMEDL. As seen, the feature activations are minimal when using CT-only but shows a clear preferential boundary activation when incorporating the MR information. Figure~\ref{fig:featuremap}(b) also shows a pseudo MR produced from a CT (Figure~\ref{fig:featuremap}(a)).

	\bumpup
	\section{Discussion}
	\bumpup
	We developed a novel approach for segmenting lung tumors located in areas with low soft-tissue contrast by leveraging learned prior information from more informative MR modality. These cross-modality priors are learned from unrelated patients and are used to hallucinate MRI to inform CT segmentation. Through extensive experiments on two different network architectures, we showed that leveraging a more informative modality (MRI) to inform inference in a less informative modality (CT), improves segmentation. Our work is limited by lack of sufficiently large MR datasets to potentially improve the accuracy of cross-domain adaptation models. Nevertheless, this is the first approach to our knowledge that used the cross-modality information in a novel way to generate CT segmentation.
	\bumpup
	\section{Conclusions}
	\bumpup
	We introduced a novel approach for segmenting on CT datasets that can leverage more informative MR modality through cross-modality learning. Our approach implemented on two different segmentation architectures shows improved performance over CT-only methods.
	\section{Acknowledgements}
	This work was supported by the MSK Cancer Center support grant/core grant P30 CA008748, and NCI R01 CA198121-03.
	
	{\tiny
		\bibliographystyle{splncs}
	}
	\bumpup
	\bibliography{refs}
	\bumpup
\end{document}